\documentclass[journal]{IEEEtran}

\usepackage{multirow}
\usepackage{array} 
\usepackage{longtable}
\usepackage{booktabs} 
\usepackage{makecell}
\usepackage{diagbox}
\usepackage{subfig}
\usepackage{amssymb}
\usepackage{graphicx}
\usepackage[flushleft]{threeparttable}
\usepackage{pifont}
\usepackage{xcolor}
\definecolor{ao}{rgb}{0.0, 0.5, 0.0}
\definecolor{amber}{rgb}{1.0, 0.49, 0.0}
\newcommand{\yestick}{{\color{ao}\ding{51}}}
\newcommand{\notick}{{\color{red}\ding{55}}}
\newcommand{\partialtick}{{\textbf{\color{amber}$\mathord{?}$}}}

\begin{document}

\title{Evaluation of Data Processing and Machine Learning Techniques in P300-based Authentication using Brain-Computer Interfaces}

\author{Eduardo López Bernal$^{1}$, Sergio López Bernal$^{*1}$, Gregorio Martínez Pérez$^{1}$, and Alberto Huertas Celdr\'an$^{2}$

\thanks{$^{*}$Corresponding author.}

\thanks{$^{1}$Eduardo López Bernal, Sergio López Bernal, and Gregorio Mart\'inez P\'erez are with the Department of Information and Communications Engineering, University of Murcia, 30100 Murcia, Spain {\tt\small (eduardo.lopez5@um.es, slopez@um.es; gregorio@um.es)}.}
\thanks{$^{2}$Alberto Huertas Celdr\'an is with the Communication Systems Group (CSG) at the Department of Informatics (IfI), University of Zurich UZH, 8050 Zürich, Switzerland {\tt\small (e-mail: huertas@ifi.uzh.ch).}}}


\maketitle


\begin{abstract}
Brain-Computer Interfaces (BCIs) are used in various application scenarios allowing direct communication between the brain and computers. Specifically, electroencephalography (EEG) is one of the most common techniques for obtaining evoked potentials resulting from external stimuli, as the P300 potential is elicited from known images. The combination of Machine Learning (ML) and P300 potentials is promising for authenticating subjects since the brain waves generated by each person when facing a particular stimulus are unique. However, existing authentication solutions do not extensively explore P300 potentials and fail when analyzing the most suitable processing and ML-based classification techniques. Thus, this work proposes i) a framework for authenticating BCI users using the P300 potential; ii) the validation of the framework on ten subjects creating an experimental scenario employing a non-invasive EEG-based BCI; and iii) the evaluation of the framework performance defining two experiments (binary and multiclass ML classification) and three testing configurations incrementally analyzing the performance of different processing techniques and the differences between classifying with epochs or statistical values. This framework achieved a performance close to 100\% f1-score in both experiments for the best classifier, highlighting its effectiveness in accurately authenticating users and demonstrating the feasibility of performing EEG-based authentication using P300 potentials.
\end{abstract}


\IEEEpeerreviewmaketitle

\section{Introduction}
\label{sec:introduction}

\IEEEPARstart{A}{}uthentication systems have traditionally relied on usernames and passwords to ensure access to restricted resources or locations. Due to the evolution of cybersecurity, new mechanisms based on tokens or biometric data, such as fingerprints \cite{Tan2019fingerprint}, voice \cite{Shenai2021voice}, behavior \cite{sanchez2021authcode}, or retinal recognition \cite{Jarina2022Retina}, have emerged in the last decades. However, these authentication systems are still vulnerable to attacks able to impersonate legitimate users \cite{Rui:limitations_auth:2019}. To improve this challenge, the literature has identified novel authentication mechanisms based on biosignals. In particular, using Brain-Computer Interfaces (BCIs) to acquire brain waves for authentication is a promising research area with much potential \cite{Rathi:auth_BCI:2021}.

BCIs are bidirectional systems able to acquire neural data or perform neurostimulation procedures. These interfaces have been traditionally employed in medicine, where the acquisition of neural data helps diagnose neurological conditions such as epilepsy \cite{Sowndhararajan2018epilepsy}. Moreover, BCIs can stimulate the brain to reduce the effects of neurodegenerative diseases, such as Parkinson's, when drug-based treatments are ineffective \cite{Khabarova2018parkinson}. Furthermore, BCIs can also be used to identify mental actions intended by the subject, like controlling a wheelchair or a robotic limb, which is helpful for neurorehabilitation \cite{Simona2018exoskeleton, Francis2021wheelchair}. Due to the evolution of BCI technologies, their use has been extended to other applications outside the medical domain. In particular, they are commonly used in video games, where players can perform actions on avatars through brain waves \cite{Minkyu2014videogames}. Finally, BCIs have gained relevance in recent years for their use in authentication since brain waves are unique to each person. Moreover, the task performed by the user to authenticate can change in case the data is compromised, being a promising alternative to traditional biometric authentication methods. 

These systems can also be classified according to their invasiveness over the brain in invasive and non-invasive technologies. The best example of a non-invasive acquisition technique is electroencephalography (EEG), which consists in obtaining the brain signals captured by electrodes placed on the scalp \cite{Lotte2015EEG}. EEG allows the acquisition of different characteristic patterns contained in brain waves, being Event-Related Potentials (ERPs) responses generated from external stimuli. One of the most relevant ERPs is the P300 potential, generated by visual or auditory stimuli \cite{Enrique2022P300}. One of the ways to elicit a P300 is by following the Oddball paradigm, based on showing a certain number of images with a frequency of approximately 250-500 ms, of which 10-20\% are images known to the subject (target), and the rest are unknown (non-target) \cite{Jang2011oddball}. This potential can be observed in the EEG signals as a positive peak around 300 ms after the stimulus, although it can vary from 250 ms to 500 ms \cite{Lotte2015EEG}. Furthermore, the use of ML classification is essential to automatically identify P300 potentials, widely used in the EEG-based authentication literature \cite{Rathi:auth_BCI:2021}. 

Concerning authentication, BCIs have been widely studied in the literature to determine whether users accessing the system are legitimate. Most of the publications focus on tasks such as being relaxed, imagining the movement of a limb, or performing a mathematical operation, among others. However, another task yet to be extensively explored in the literature is to authenticate users based on P300 potentials using the Oddball paradigm. Moreover, there is no consensus in the literature on which processing or classification techniques are better to solve this problem, thus generating an opportunity currently unexplored. Focusing on ML classification, the literature typically relies on a binary approach, although many works do not specify what type of classification they use. The related work also focuses only on classification based on epochs, while the use of statistical values extracted from those epochs is unexplored. Finally, there is an open challenge to determine which sliding window size is better for solving statistics-based classification in terms of vectors, where the present work defines as an hypothesis that a larger windows size would offer better results. Based on the above, there is a need for systems capable of performing the authentication process autonomously and transparently.

Intending to improve the previous limitations, the present publication provides the following contributions:
\begin{itemize}
    \item The design and implementation of a framework managing all the phases of the BCI cycle to authenticate subjects through visual P300 potentials contained in brain waves, available in \cite{Lopez:Auth-Code:2023}. In particular, this framework can synchronize EEG signals with the stimuli presented, implementing different EEG data processing techniques. Moreover, it provides both binary and multiclass classification approaches, allowing training using epochs or statistical values from the epochs. Finally, it can evaluate different window sizes of vectors during statistics-based classification.
    \item The validation of the proposed framework on ten subjects in a experimental scenario using a non-invasive EEG-based BCI capable of acquiring P300 potentials generated following the Oddball paradigm. In particular, the subjects underwent ten experiment divided in two days, where each experiment consists in the presentation of known and unknown visual stimuli to elicit P300 potentials. 
    \item The evaluation of the framework performance through two different experiments, testing for each of them three different configurations of processing and classification techniques based on the EEG data acquired by the framework. The first experiment evaluates a binary ML classification approach, while the second focuses on a ML multiclass perspective. Regarding the testing configurations, the first one uses the Notch and Butterworth filters, performing the classification using epochs. In contrast, the second considers adding ICA to the processes defined by the first configuration. Finally, the third configuration is cumulative to the previous two. It uses ML classification through statistical values extracted from the epochs using four window sizes. The best-performing classifier for both binary and multiclass approaches obtains an f1-score close to 100\%. 
\end{itemize}

The remainder of this article is organized as follows. Section~\ref{sec:related} reviews the literature performing the authentication process using P300 potentials. Section~\ref{sec:scenario} presents the scenario designed to perform the experiments, while Section~\ref{sec:framework} describes each module included in the proposed framework to authenticate subjects. After that, Section~\ref{sec:experiments} shows the experiments defined and the results obtained for each of the three configurations tested. Furthermore, Section~\ref{sec:discussion} discusses the results provided by this work, also comparing them with the results from the literature. Finally, Section~\ref{sec:conclusions} presents some conclusions and future work.

\section{Related work}
\label{sec:related}

This section analyzes the literature focused on subject authentication, using the Oddball paradigm to generate the P300 evoked potential in brain waves, presenting different types of visual stimuli to subjects. 

In 2012, Gupta et al. \cite{Gupta:auth_BCI:2012} studied three variations of the Oddball paradigm in eight subjects, using an eight-channel BCI with a sampling rate of 256 Hz. This work performed two trials per variation, where each trial consisted in flashing a letter in a mental spelling scenario, ranging between 40 and 48 times. Finally, the subjects indicated the number of times the target letter was highlighted. The authors filtered the data using a forward-reverse Butterworth filter with cutoff frequencies of 1-12 Hz, later normalizing the data. Finally, this publication employed triple cross-validation using Bayesian Linear Discriminant Analysis (LDA), obtaining 90\% accuracy.

Yu et al. \cite{Yu:auth_BCI:2014}, in 2014, proposed an authentication system that displayed images of different people on the screen, using a BCI with eight channels and a sampling rate of 250 Hz, in which four subjects participated. The authors obtained epochs from the stimulus onset time to 800 ms post-stimulus presentation. The experiment applied different processing techniques, such as a Butterworth filter with 1-12 Hz cutoff frequencies, decimation, winsorizing, and standardization. Moreover, this work used a binary P300 classification model using ten-fold cross-validation and Fisher’s Linear Discriminant Analysis (FLDA) to authenticate. This model obtained 83.1\% accuracy with sensitivity and specificity equal to 0.570 and 0.897, respectively.

In 2016, Koike-Akino et al. \cite{Koike-akino:auth_BCI:2016} conducted an experiment in which each subject selected one card from the five that appeared on the screen. Specifically, 25 subjects participated, using a BCI with 14 channels and a sampling rate of 128 Hz. The researchers obtained epochs from 100 ms pre-stimulus to 700 ms post-stimulus presentation, using Principal Component Analysis (PCA) and Partial Least Squares (PLS) to reduce dimensionality. Furthermore, as classifiers, this solution used LDA, Quadratic Discriminant Analysis (QDA), Naive Bayes (NB), Decision Trees (DT), k-Nearest Neighbors (k-NN), Support Vector Machine (SVM), Logistic Regression (LR), and Deep Neural Networks (DNN), all of them following a binary classification approach. LDA and PLS with 16 epochs achieved the best performance with an accuracy of 96.7\%.

In 2019, Zeng et al. \cite{Zeng:auth_BCI:2019} proposed an authentication system with two different attack scenarios in which 45 subjects participated using a BCI with 16 channels and 2400 Hz as the sampling rate. In the first scenario, a subject considered an impostor visualized the sequence of images of the legitimate subject without knowing what the target image was. In contrast, in the second scenario, the impostor knew the target image. The authors combined the data from both experiments and filtered the raw data by a low-pass Chebyshev digital filter with a band-pass of 40 Hz and a stop-band of 49 Hz. After that, the processing phase downsampled from 2400 Hz to 600 Hz, subsequently extracting epochs from 200 ms before the stimulus to 1000 ms after the event. Finally, the ML algorithm offering the best performance was the combination of Hierarchical Discriminant Component Analysis (HDCA) and Genetic Algorithm (GA), obtaining an accuracy of 94.26\%.

Kaongoen et al. \cite{Kaongoen:auth_BCI:2020}, in 2020, made an extension of their previous work \cite{Yu:auth_BCI:2014}, where a total of ten subjects participated. This work mixed a small number of target stimuli with a large number of non-target stimuli, then using a random oversampling method to balance both P300 and non-P300 samples. This research used FLDA to build the P300 classification models, testing its performance when an external attacker did not have access to the target image used (veiled attack) and when it was known (unveiled attack), offering for both an FFR equal to zero. The veiled attack obtained a FAR of 0.003, while the unveiled attack reached a 0.010 value.

Rathi et al. \cite{Rathi:auth_BCI:2021} conducted a study in 2021 in which ten participants had to view a sequence of images and then blink for five seconds. This work used a BCI with ten channels and a sampling rate of 512 Hz. During the last ten seconds of the experimentation, the subjects had to be at rest. These experiments processed the data with a sixth-order Butterworth band-pass filter between 0.1 and 30 Hz. After that, they used PCA and the Chebyshev Type 1 low-pass filter, then normalizing the data. The authors compared the EEG obtained in the verification phase with previous recordings. If the match was greater than a threshold, the subject was accepted; otherwise, the access was denied. The best algorithm to authenticate subjects was QDA, obtaining 97\% accuracy.

Finally, \tablename~\ref{table:literature} summarizes the previous publications, comparing the present work with the existing literature. This table highlights that no work in the literature uses either the Notch filter or ICA. In contrast, all of them except two use the Butterworth filter. Regarding data classification, these works only use binary classification, although four do not indicate the classification approach followed. Finally, five of the seven works perform the classification by epochs, while non of them employ statistical values. Focusing on the present work, ten subjects participated, employing a BCI with eight channels and a sampling rate of 256 Hz since these numbers align with the existing literature. Likewise, the classification is performed using epochs as the related work highlights that excellent results can be obtained. However, as current literature does not consider obtaining new features on the original data that could better represent the particularities of the EEG and the P300, this work also explores the possibility of using a classification based on statistical values extracted from the epochs. Finally, the work at hand also considers a multiclass approach to identify to which user the brain waves belong.

\begin{table*}[!htb]
\centering
\caption{Comparison of the different works analyzed from the literature.}
\label{table:literature}
\begin{tabular}{@{}lccccccccccc@{}}
\toprule
Reference & Year & Subjects & Electrodes & Sampling rate & \begin{tabular}{@{}c@{}}Notch \\ filter\end{tabular} & \begin{tabular}{@{}c@{}}Butterworth \\ filter\end{tabular} & ICA & \begin{tabular}{@{}c@{}}Binary \\ classification\end{tabular} & \begin{tabular}{@{}c@{}}Multiclass \\ classification\end{tabular} & Epochs & Statistics\\
\midrule
\cite{Gupta:auth_BCI:2012} & 2012 & 
\makecell{8} & \makecell{8} & \makecell{256 Hz} & \notick & \yestick & \notick & \partialtick & \partialtick &
\partialtick &
\notick
\\ \midrule

\cite{Yu:auth_BCI:2014} & 2014 & 
\makecell{4} & \makecell{8} & \makecell{250 Hz} & \notick & \yestick & \notick & \yestick & \notick &
\yestick &
\notick
\\ \midrule

\cite{Koike-akino:auth_BCI:2016} & 2016 & 
\makecell{25} & \makecell{14} & \makecell{128 Hz} & \notick & \notick & \notick & \notick & \yestick &
\yestick &
\notick
\\ \midrule

\cite{Zeng:auth_BCI:2019} & 2019 &
\makecell{45} & \makecell{16} & \makecell{2400 Hz} & \notick & \notick & \notick & \partialtick & \partialtick &
\yestick &
\notick
\\ \midrule

\cite{Kaongoen:auth_BCI:2020} & 2020 &
\makecell{15} & \makecell{8} & \makecell{250 Hz} & \notick & \yestick & \notick & \yestick & \notick &
\yestick &
\notick
\\ \midrule

\cite{Rathi:auth_BCI:2021} & 2021 &
\makecell{10} & \makecell{10} & \makecell{512 Hz} & \notick & \yestick & \notick & \partialtick & \partialtick &
\yestick &
\notick
\\ \midrule

This work & 2023 & 
\makecell{10} & \makecell{8} & \makecell{256 Hz} & \yestick & \yestick & \yestick & \yestick & \yestick &
\yestick &
\yestick \\ \midrule
\end{tabular}
\begin{tablenotes}
\item \yestick\space used,\space\partialtick\space not indicated,\space\notick\space not used by the work
\end{tablenotes}
\end{table*}

\section{Scenario}
\label{sec:scenario}

\begin{figure}[ht]
\includegraphics[scale=0.90]{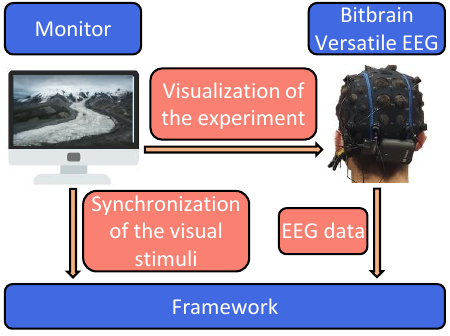}
\centering
\caption{Components integrating the implemented solution.}
\label{fig:scenario}
\end{figure}

This section presents an experimental scenario in a company composed of ten workers aiming to authenticate their users. For that, \figurename~\ref{fig:scenario} details the components and actors involved in the implemented scenario, such as a monitor to present the visual stimuli following the Oddball paradigm and a BCI to acquire the EEG signals produced by the subjects while visualizing the images. The last component of the solution is a framework in charge of receiving the signal obtained by the BCI and synchronizing the visual stimuli to know their type (target or non-target) and the instant each stimulus is shown to the subject. Furthermore, the framework performs the necessary tasks to process the data, detect the P300 contained in the signals, and authenticate the subjects based on them.

\subsection{Participants}
This study evaluated ten healthy subjects without any known neurological disease, particularly six men and four women, aged between 23 and 55. The experiments recorded brain waves from each subject while visualizing the stimuli presented. In total, the subjects underwent 20 tests on two different days, performing ten on each day. Before starting the first session, the participants were told to be upright and perpendicular to the ground, preventing movements that could introduce noise to the acquired signals. In addition, the users received detailed information about the purpose and procedure of the experimentation.

\subsection{Visual stimuli}
The stimuli presented to the subjects consisted in visualizing a video while the EEG was monitored. The video contained static images shown sequentially over time following the Oddball paradigm, containing images from two types: target and non-target. Target images were known to each subject, while non-target stimuli were unknown. For each of the 20 experiments performed and each subject, the software randomly selected stimuli from both categories without any meaning for the subjects, being able to repeat images between subjects. In addition, the stimuli within each video were randomly ordered, so the target image was displayed throughout the experiment in arbitrary positions.

Each experiment presented the target image at the beginning of the video for five seconds, repeating it 40 times throughout the experiment in random positions. In addition, each experiment showed 160 unknown images, thus creating a video with a total of 200 images. Each image was displayed for one second, so the experiment duration was 205 seconds. Finally, the data acquisition had a total duration of 213 seconds to ensure the correct recording of all samples.

\subsection{BCI}
The BCI device employed in the experiments was the eight-channel Bitbrain Versatile EEG that uses semi-dry electrodes and has a sampling rate of 256 Hz. For the generation of the EEG signal, the experiments used the eight electrodes available, placing them in positions Fp1, Fp2, C3, C4, P7, P8, O1, and O2 following the international 10-20 system \cite{Homan1987system10-20}. This work chose these positions since they are widely used in the literature due to their coverage of the parieto-occipital area of the brain, where cerebral activity related to vision can be acquired \cite{Takano2014OccipitalArea}.

\section{Framework}
\label{sec:framework}

This section describes the design and implementation in Python of a framework representing the phases of the BCI operational cycle: data acquisition, data processing, feature extraction, and classification \cite{Lopez2020Survey}. Specifically, this framework is based on the implementation provided by Mart\'inez et al. \cite{Enrique2022P300}. The code of the framework is publicly available at Github \cite{Lopez:Auth-Code:2023}. Furthermore, this section presents the creation of authentication datasets depending on whether the classification employs epochs or statistical values or whether the framework uses a binary or multiclass classification.

\begin{figure}[ht]
\includegraphics[width=\columnwidth]{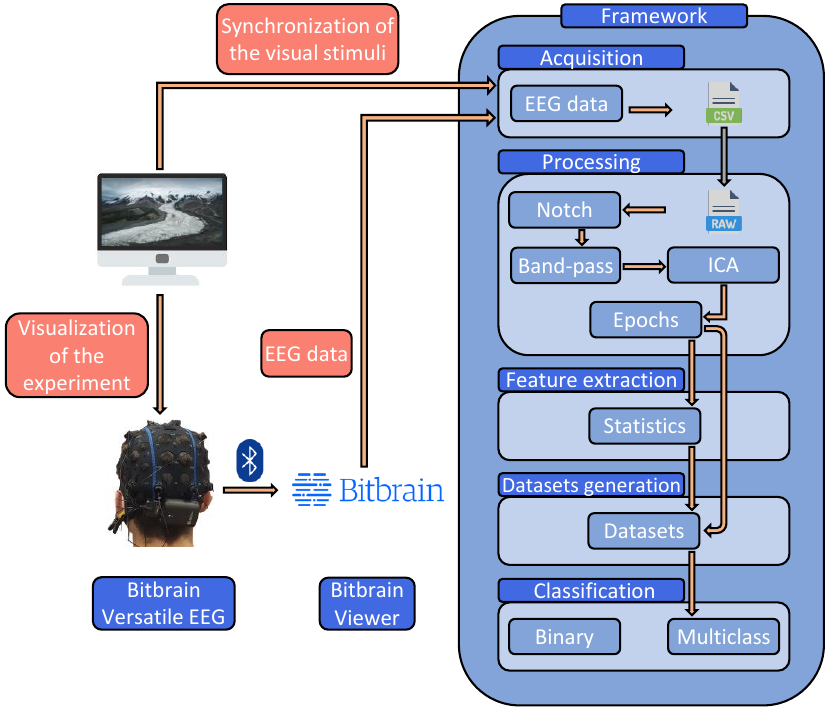}
\centering
\caption{Conceptual design of the proposed framework.}
\label{fig:framework}
\end{figure}

\figurename~\ref{fig:framework} shows the conceptual design of the framework, whose modules and components are presented in more detail in the following sections. The data coming from the BCI are first transmitted to the acquisition module, which synchronizes the signals with the stimuli presented to the user, subsequently storing the data in a CSV file. These data are then used by the processing module, which aims to improve the quality of the EEG signals, being able to apply Notch, Butterworth, and ICA techniques. Furthermore, this software block allows to extract epochs from the data. Once the processing tasks are completed, it is possible to generate datasets directly or perform an intermediate feature extraction process. In the former, the framework composes the authentication datasets that serve as input to the classification block. In the latter, the framework extracts representative statistical values of the epochs generated in the processing module. Once these stages are completed, the classification module performs the authentication of the users based on the data provided by the previous modules. Specifically, this block uses different classification algorithms, which are computational learning elements that offer a prediction based on previous training. In particular, the framework implements both binary and multiclass approaches.

\subsection{Data acquisition}
\label{sec:acquisition}

The acquisition of the EEG signal is performed through the BCI electrodes that are in contact with the scalp. The Bitbrain Viewer software, which is in charge of communicating with the BCI, allows for establishing a Lab Streaming Layer (LSL) communication to retrieve data from the BCI controller. The framework, to obtain the EEG signal transmitted through LSL, creates an independent thread that remains listening for any EEG data stream.

This module obtains the EEG data from the BCI and synchronizes the visual stimuli displayed on the screen. The use of the Oddball paradigm requires to know the instant each stimulus is shown to the subject and its type (target or non-target) to subsequently study the generated P300 waves. For that, this module adds in a CSV file the timestamp in milliseconds for each sample obtained and the EEG values captured in each channel for each time instant, also indicating the type of stimulus presented to the subject. 

\subsection{Data processing}
\label{sec:processing}

This module applies different signal processing techniques to obtain the cleanest possible signal to authenticate. It comprises four components (Notch, Butterworth, ICA, and epochs) implemented using the Python MNE library, explained below in detail.

Before applying data processing, this module imports the data obtained in the acquisition phase. Then, the first component of the processing stage is the Notch filter, used to attenuate the frequency at 50 Hz due to the noise produced by the wiring of the European electrical system. The second component is a sixth-order Butterworth band-pass filter, useful to remove unwanted
frequency effects and biases. Different experiments in this work concluded that a fourth-order filter, widely used in the literature, worsened the authentication results. Moreover, the cut-off frequency was set at 1-17 Hz, for which different tests were performed to find the best range, consistent with the literature studied.

To complement the filters used, the framework introduces a component implementing ICA, which is used to suppress signal artifacts persisting after applying the previous filters, such as eye blinking. The next component implemented consists in generating epochs on the data, selecting 0.1 seconds before the event occurs and 0.8 seconds afterward. This range prevents EEG signals belonging to two images from overlapping, which aligns with the ranges defined in the literature. Finally, the framework stores the epochs into a CSV file, later used for subsequent modules. Once the epochs are created, the framework allows performing the classification either by using those epochs or statistical values extracted from the epochs. Thus, the framework selects the adequate module to continue the execution, based on the category of classification desired.

Other techniques used in the literature using the Oddball paradigm and P300 potentials were also tested in the framework, finally discarded because they did not offer good results for authentication. First, the authors tested the downsampling technique to reduce the number of signal samples per second. In particular, this paper used a reduction to 50 Hz since it is the most common sample size in works using P300. However, this filter was discarded since the results obtained were similar to those not applying it, thus providing more samples for the classification phase. This research also considered the application of the Windsorising technique, calculating the 10th and 90th percentiles for each EEG channel. This filter was excluded from the framework as it modified the voltages of all the EEG channels except Fp1 and Fp2, affecting the P300 wave and, thus, the classification performance. Finally, the authors considered PCA and decimation to reduce the feature vector size but, since the number of features was reduced, this technique did not positively influence the classification.

\subsection{Feature extraction}
\label{sec:featureExtraction}

This module is in charge of extracting statistical values from the epochs created in the previous stage. The process first consists in reading the CSV file containing the subject's target epochs, later using a sliding window with four different values to obtain the statistics. In particular, the window sizes are 58, 116, 174, and 232, where these values correspond to different fractions (25\%, 50\%, 75\%, and 100\%) of the total of vectors in an epoch, which are 232. The statistical values extracted through the sliding window are mean, variance, standard deviation, maximum, sum, and median. Finally, the values extracted are stored in a CSV file, where each row contains the statistical values obtained at each iteration. 

\subsection{Datasets generation}
\label{sec:datasets}

This module creates the authentication datasets used in the classification module to determine if the subjects are legitimate employees of a company. These datasets are generated depending on the approach used in the classification module (i.e. binary or multiclass approach). In binary classification, a dataset must be created per user, while in a multiclass approach, a common dataset is required for all the subjects.

Focusing on binary classification, and assuming the use of epochs instead of statistical values, obtaining the dataset for each subject first consists in reading the CSV file containing the epochs of that particular subject generated in the processing module. Subsequently, those epochs labeled as non-target are removed. Finally, the framework creates a dataset containing the target epochs of the subject being authenticated and a set of target epochs from the rest of the subjects (i.e., illegitimate), labeled as non-target. To do this, the number of target epochs of the subject is divided by the number of the remaining subjects (nine), aiming to find how many target epochs need to be extracted from each subject. For each illegitimate subject, the module acquires random epochs from the target epoch set, then added to the dataset of the legitimate subject labeled as non-target. The module selects the same number of target and non-target epochs to have a balanced dataset. In contrast, by assuming statistical values, the framework creates a dataset following the same steps presented above using the CSV files containing statistical values. For both approaches, the dataset generation process is performed five times for each subject to ensure data variability and, thus, more reliable results. In conclusion, this module generates five datasets for each subject, that is, 50 datasets in total per approach. 

In contrast, the framework uses for multiclass classification a balanced dataset with the same epochs for each subject. To do this, the framework loads the files containing the epochs of each subject and, as in binary classification, the epochs labeled as non-target are eliminated. After that, the framework balances the samples between the subjects by selecting the user with the least number of epochs, randomly removing the samples from the other subjects so that they have the same number. The selected epochs of each subject are concatenated to form a set of epochs common to all users, preserving to which user each epoch belongs to. As in binary classification, the same steps are performed when working with statistical values, but working with vectors instead of epochs. Furthermore, this classification category generates five datasets to obtain more reliable results.

The authors evaluated different techniques applied to every dataset in both dataset generation strategies, such as data normalization, standardization, or PCA. However, none of these techniques offered better results than the previous configurations. Because of that, these techniques were not included in the framework.

\subsection{Classification}
\label{sec:classification}

Once the framework creates the datasets explained in the previous section, the authentication process starts. Both binary and multiclass ML classification can be performed to evaluate which approach is the most appropriate for the authentication process. This analysis is essential since the literature does not compare in detail these approaches and could help provide classification systems with better performance.

The framework uses several ML algorithms, such as Logistic Regression (LR) with default hyperparameters. Another algorithmn used is Linear Discriminant Analysis (LDA) with ``auto'' as shrinkage and ``eigen'' as the solver, or Minimum Distance to Mean (MDM) with default hyperparameters. Furthermore, this module uses Random Forest (RF) with 42 as random\_state, or Quadratic Discriminant Analysis (QDA) by default hyperparameters. The two last algorithms used are Support Vector Machine (SVM) with ``scale'' as gamma and K-Nearest Neighbors (kNN) with 50 n\_neighbors. Furthermore, several configurations have been created using the above algorithms. In particular, the framework includes Vectorizer (Vect), StandardScaler (SS), and TangentSpace (TS) with default hyperparameters. In addition, XDawn uses 2 as nfilter and 1 as classes, while ERPCovariance (ERPC) employs ``oas'' as the estimator. Based on the combination between the previous algorithms and techniques, the framework implements 21 classifiers. \tablename ~ \ref{table:classifiers} shows the configurations for each trained classifier in the experimentation phase.

Finally, the framework extracts the classification report for each classifier, thus obtaining the main classification metrics, such as accuracy, precision, recall, and f1-score. In addition, it obtains confusion matrices to evaluate the performance of the classification.

\begin{table}[!htb]
\centering
\caption{Configurations of different trained classifiers resulting from combining different techniques with ML algorithms.}
\label{table:classifiers}
\resizebox{\columnwidth}{!}{
\begin{tabular}{@{}lcccccccccccc@{}}
\toprule
Cls & \multicolumn{5}{c}{Techniques} & \multicolumn{7}{c}{Algorithms}

\\
\cmidrule(lr){2-6}
\cmidrule(lr){7-13}
&

Vect & SS & XDawn & ERPC & TS & LR & LDA & MDM & RF & QDA & SVM & KNN\\
\midrule
Cl 1 &
\makecell{\checkmark} & \makecell{\checkmark} & \makecell{} & \makecell{} & \makecell{} & \makecell{\checkmark} & \makecell{} & \makecell{} & \makecell{} & \makecell{} & \makecell{} & \makecell{}
\\ \midrule

Cl 2 &
\makecell{\checkmark} & \makecell{} & \makecell{} & \makecell{} & \makecell{} & \makecell{} & \makecell{\checkmark} & \makecell{} & \makecell{} & \makecell{} & \makecell{} & \makecell{}
\\ \midrule

Cl 3 &
\makecell{\checkmark} & \makecell{} & \makecell{\checkmark} & \makecell{} & \makecell{} & \makecell{} & \makecell{\checkmark} & \makecell{} & \makecell{} & \makecell{} & \makecell{} & \makecell{}
\\ \midrule 

Cl 4 &
\makecell{} & \makecell{} & \makecell{} & \makecell{\checkmark} & \makecell{\checkmark} & \makecell{\checkmark} & \makecell{} & \makecell{} & \makecell{} & \makecell{} & \makecell{} & \makecell{}
\\ \midrule 

Cl 5 &
\makecell{} & \makecell{} & \makecell{} & \makecell{\checkmark} & \makecell{} & \makecell{} & \makecell{} & \makecell{\checkmark} & \makecell{} & \makecell{} & \makecell{} & \makecell{}
\\ \midrule

Cl 6 &
\makecell{\checkmark} & \makecell{} & \makecell{} & \makecell{} & \makecell{} & \makecell{} & \makecell{} & \makecell{} & \makecell{\checkmark} & \makecell{} & \makecell{} & \makecell{}
\\ \midrule

Cl 7 &
\makecell{\checkmark} & \makecell{} & \makecell{} & \makecell{} & \makecell{} & \makecell{} & \makecell{} & \makecell{} & \makecell{} & \makecell{\checkmark} & \makecell{} & \makecell{}
\\ \midrule 

Cl 8 &
\makecell{\checkmark} & \makecell{} & \makecell{} & \makecell{} & \makecell{} & \makecell{} & \makecell{} & \makecell{} & \makecell{} & \makecell{} & \makecell{\checkmark} & \makecell{}
\\ \midrule 

Cl 9 &
\makecell{\checkmark} & \makecell{} & \makecell{} & \makecell{} & \makecell{} & \makecell{} & \makecell{} & \makecell{} & \makecell{} & \makecell{} & \makecell{} & \makecell{\checkmark}
\\ \midrule

Cl 10 &
\makecell{\checkmark} & \makecell{} & \makecell{\checkmark} & \makecell{} & \makecell{} & \makecell{} & \makecell{} & \makecell{} & \makecell{\checkmark} & \makecell{} & \makecell{} & \makecell{}
\\ \midrule

Cl 11 &
\makecell{} & \makecell{} & \makecell{} & \makecell{\checkmark} & \makecell{\checkmark} & \makecell{} & \makecell{} & \makecell{} & \makecell{\checkmark} & \makecell{} & \makecell{} & \makecell{}
\\\midrule

Cl 12 &
\makecell{\checkmark} & \makecell{} & \makecell{} & \makecell{\checkmark} & \makecell{} & \makecell{} & \makecell{} & \makecell{} & \makecell{\checkmark} & \makecell{} & \makecell{} & \makecell{}
\\ \midrule

Cl 13 &
\makecell{\checkmark} & \makecell{} & \makecell{\checkmark} & \makecell{} & \makecell{} & \makecell{} & \makecell{} & \makecell{} & \makecell{} & \makecell{\checkmark} & \makecell{} & \makecell{}
\\ \midrule

Cl 14 &
\makecell{} & \makecell{} & \makecell{} & \makecell{\checkmark} & \makecell{\checkmark} & \makecell{} & \makecell{} & \makecell{} & \makecell{} & \makecell{\checkmark} & \makecell{} & \makecell{}
\\ \midrule

Cl 15 &
\makecell{\checkmark} & \makecell{} & \makecell{} & \makecell{\checkmark} & \makecell{} & \makecell{} & \makecell{} & \makecell{} & \makecell{} & \makecell{\checkmark} & \makecell{} & \makecell{}
\\ \midrule

Cl 16 &
\makecell{\checkmark} & \makecell{} & \makecell{\checkmark} & \makecell{} & \makecell{} & \makecell{} & \makecell{} & \makecell{} & \makecell{} & \makecell{} & \makecell{\checkmark} & \makecell{}
\\\midrule 

Cl 17 &
\makecell{} & \makecell{} & \makecell{} & \makecell{\checkmark} & \makecell{\checkmark} & \makecell{} & \makecell{} & \makecell{} & \makecell{} & \makecell{} & \makecell{\checkmark} & \makecell{}
\\ \midrule

Cl 18 &
\makecell{\checkmark} & \makecell{} & \makecell{} & \makecell{\checkmark} & \makecell{} & \makecell{} & \makecell{} & \makecell{} & \makecell{} & \makecell{} & \makecell{\checkmark} & \makecell{}
\\ \midrule

Cl 19 &
\makecell{\checkmark} & \makecell{} & \makecell{\checkmark} & \makecell{} & \makecell{} & \makecell{} & \makecell{} & \makecell{} & \makecell{} & \makecell{} & \makecell{} & \makecell{\checkmark}
\\ \midrule

Cl 20 &
\makecell{} & \makecell{} & \makecell{} & \makecell{\checkmark} & \makecell{\checkmark} & \makecell{} & \makecell{} & \makecell{} & \makecell{} & \makecell{} & \makecell{} & \makecell{\checkmark}
\\ \midrule

Cl 21 &
\makecell{\checkmark} & \makecell{} & \makecell{} & \makecell{\checkmark} & \makecell{} & \makecell{} & \makecell{} & \makecell{} & \makecell{} & \makecell{} & \makecell{} & \makecell{\checkmark}
\\ 

\bottomrule
\end{tabular}}
\end{table}

\section{Experiments and results}
\label{sec:experiments}

This section presents the definition of two experiments used to validate the framework proposed, and three different configurations per experiment, following an incremental procedure. Their particularities are subsequently presented, also highlighting the results obtained. In particular, the first experiment uses a binary classification, while the second uses a multiclass approach. It is important to note that all the results presented in this section correspond to the f1-score since it represents a more robust metric compared to accuracy. Additionally, the dataset is divided into 80\% for training and 20\% for testing in both experiments.

\subsection{Experiment 1: Binary classification}

This experiment uses binary classification to differentiate a subject from the rest based on their brain waves. The first configuration, called the base case, corresponds to applying a subset of techniques explained in Section~\ref{sec:processing}. Specifically, it uses the Notch filter to attenuate the frequency at 50 Hz, and the sixth-order Butterworth band-pass filter with cut-off frequencies of 1-17 Hz. After their application, the framework generates both the epochs and the authentication dataset to perform the classification process. The second configuration is an extension of the previous one, applying ICA just before creating the epochs, thus eliminating some artifacts that could persist after the filters. The last configuration implemented in this experiment consists in obtaining statistical values of each channel from the generated epochs, as presented in Section~\ref{sec:featureExtraction}, using these statistical values for the classification. The results are compared between configurations to determine whether using epochs or statistical values offers better performance. Finally, it is relevant to indicate that the features used to train the models are the EEG voltages obtained by each of the eight channels.

\begin{figure}[ht]
\includegraphics[width=\columnwidth]{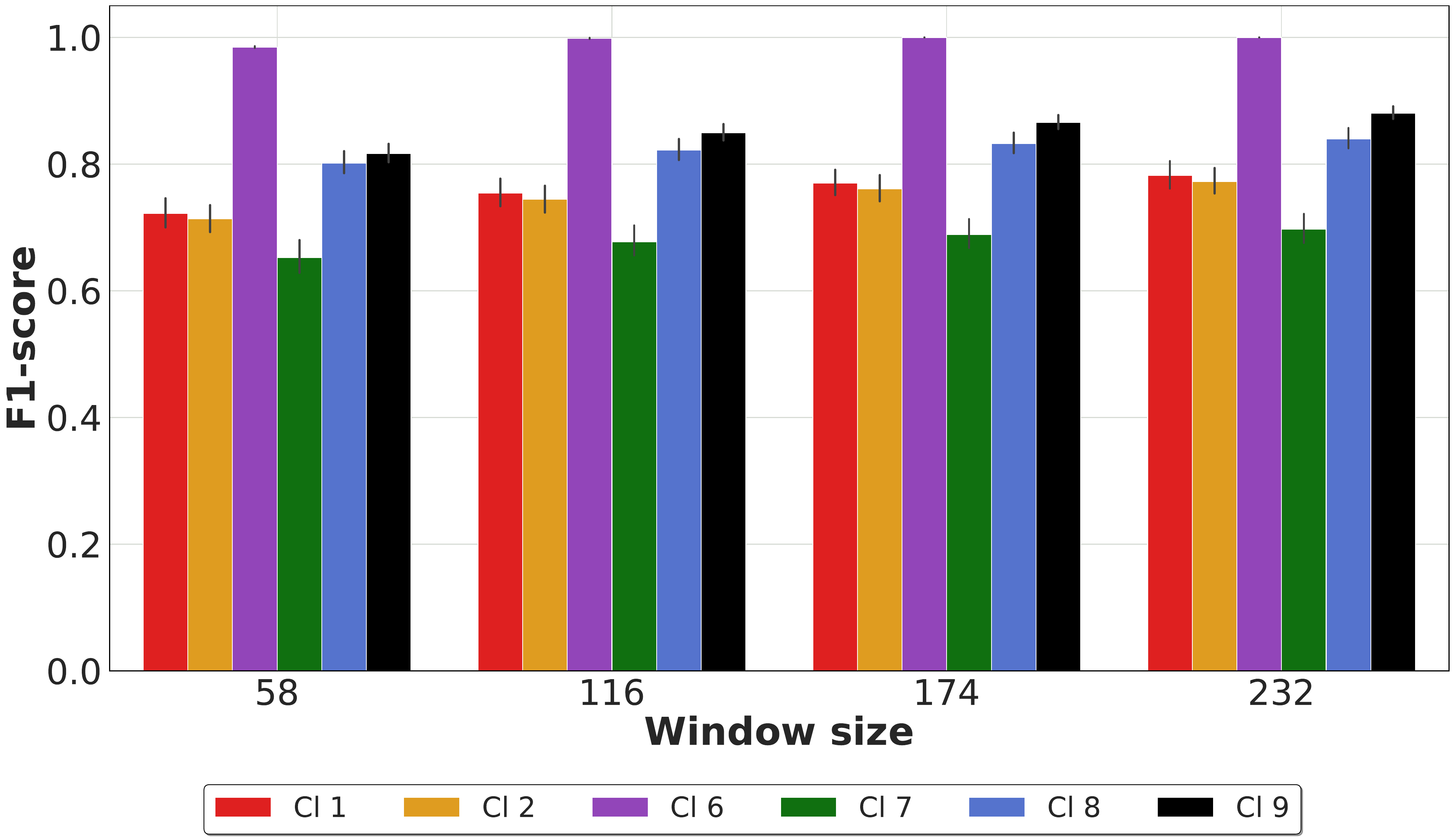}
\centering
\caption{Results obtained in the third experiment after the application of the four window sizes in the binary classification.}
\label{fig:binaryStatistics}
\end{figure}

Before comparing the results obtained between the three configurations, it is essential to present the analysis of the four different sliding window sizes for the third configuration, already presented in Section~\ref{sec:featureExtraction}. In particular, it can be observed in \figurename~\ref{fig:binaryStatistics} that only six out of 21 classifiers are represented. This reduction is motivated by using the XDawn and ERPCovariance techniques in the excluded classifiers, as their fit method requires receiving a three-dimensional array as parameter. However, the framework used a two-dimensional array, where the values of each classifier represent the mean obtained from that classifier among the datasets of all subjects. The results from \figurename~\ref{fig:binaryStatistics} indicate that the classifiers increase their performance as the window size augments by containing more information in each vector, 232 being the most promising value for this parameter.

\begin{figure}[ht]
\includegraphics[width=\columnwidth]{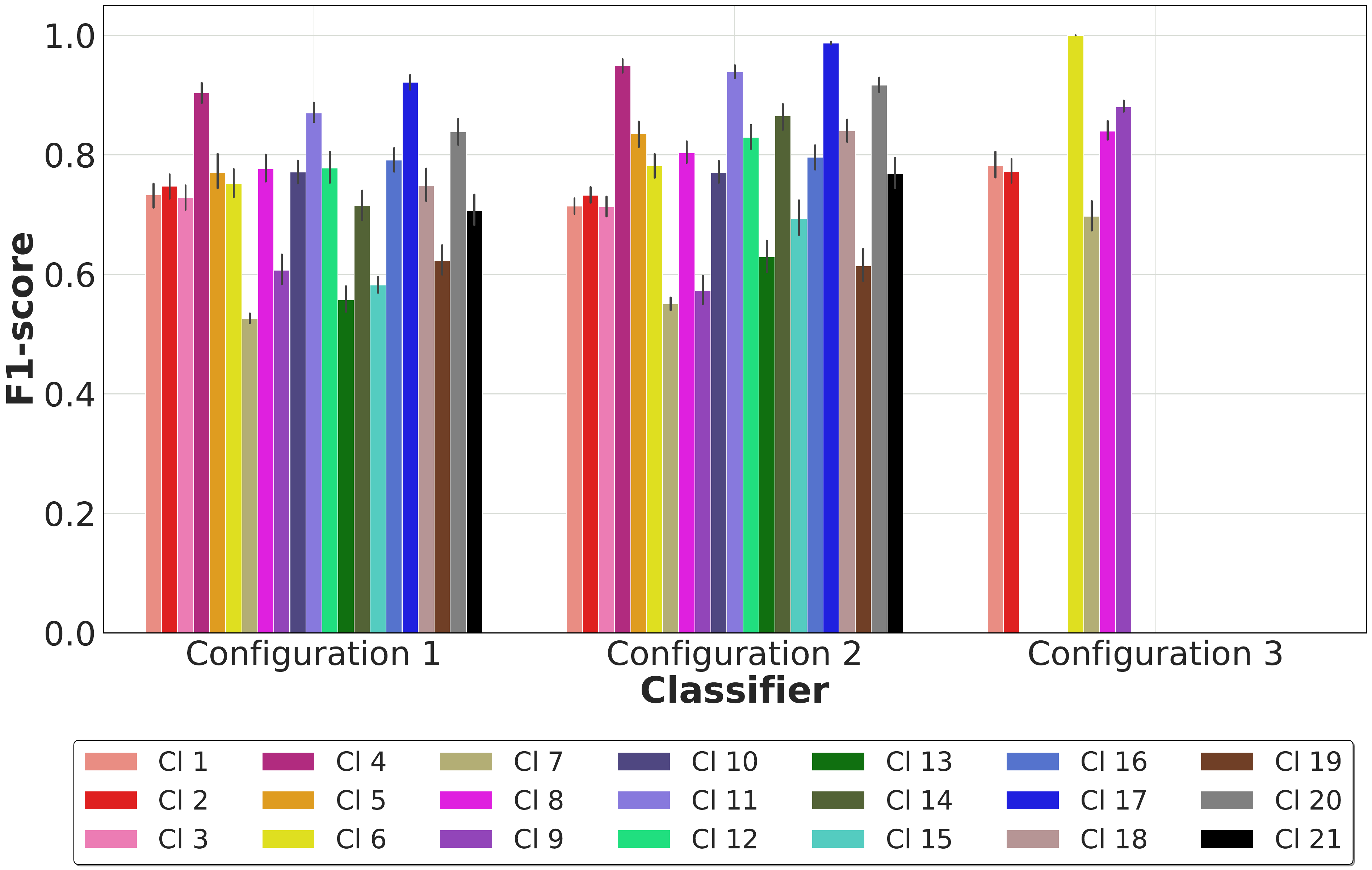}
\centering
\caption{Results obtained from the three configurations after performing a binary classification.}
\label{fig:binary}
\end{figure}

After presenting the analysis with different sliding window sizes, \figurename~\ref{fig:binary} presents the performance of the different classifiers evaluated in terms of f1-score, clustered by each of the three configurations explained above. In particular, it represents the base case, the application of ICA, and the use of statistical values through a window size equal to 232. 

In the first configuration, it can be seen how most of the classifiers produce results around 0.70 f1-score. However, classifiers 7, 13, and 15 are below this average, over 0.55. It is important to highlight that the numbering of these classifiers has already been presented in \tablename~\ref{table:classifiers}. In contrast, \figurename~\ref{fig:binary} shows that classifiers 4 and 17 obtain excellent results, reaching 0.90 and 0.92, respectively. 

After performing the second configuration, that is, after adding ICA to Notch and Butterworth filters, the results indicate a clear improvement in the results of most classifiers. Some classifiers, such as 1, 2, 3, 9, and 19, worsen by 0.01 or 0.02 compared to the base case results. Nevertheless, the application of ICA significantly improves the performance of the rest of the classifiers. For example, classifier 20 goes from 0.84 in the base case to 0.92 f1-score after applying ICA.

Finally, applying the third configuration improves the results of all the classifiers. The classifier that made the most significant increase in performance was classifier 9, going from 0.57 using epochs to 0.88 employing statistical values. Furthermore, classifier 6 obtained a performance of 1.0 for all subjects. These results indicate that employing statistical values produces a notable improvement compared to the use of epochs.

\begin{figure}[ht]
\includegraphics[width=\columnwidth]{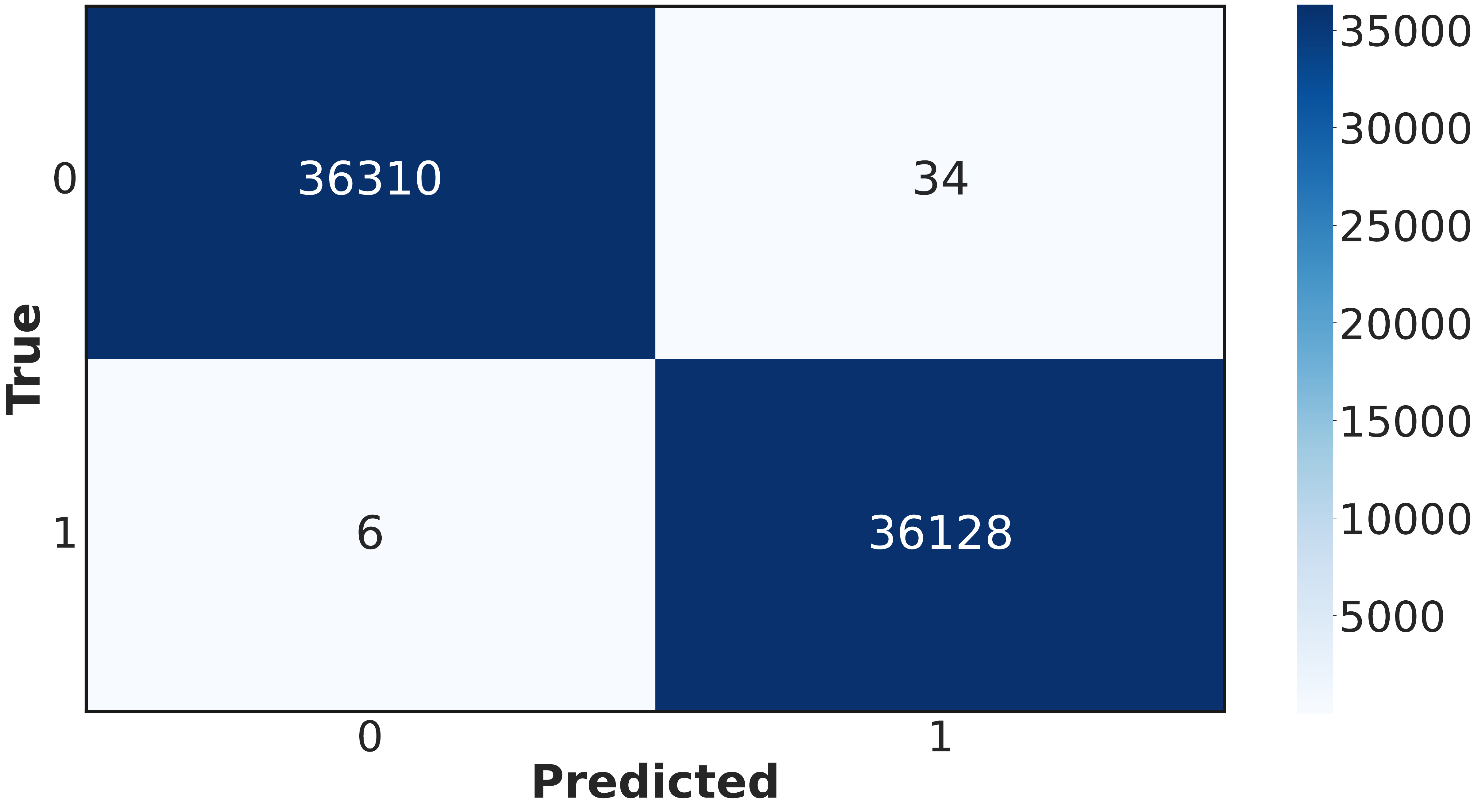}
\centering
\caption{Confusion matrix obtained for the best classifier from binary classification (classifier 6).}
\label{fig:confusionMatrixBinary}
\end{figure}

\figurename~\ref{fig:confusionMatrixBinary} shows the confusion matrix obtained for the best classifier (classifier 6) after performing the binary classification using statistical values in one random subject, indicating that most of the vectors are correctly classified. In this classifier, the training data size is 289 908 vectors, while the test data size is 72 478. Furthermore, training the model needs three minutes and 34 seconds, and 92.6 MB for model storage, while its evaluation requires one second. At this point, it is essential to indicate the hardware specifications of the machine running the framework. Particularly, it uses an Intel Core i7-5930K CPU at 3.5 GHz, with six cores and two threads per core. Moreover, the machine has installed a RAM of 94 Gb.

\subsection{Experiment 2: Multiclass classification}

\begin{figure}[ht]
\includegraphics[width=\columnwidth]{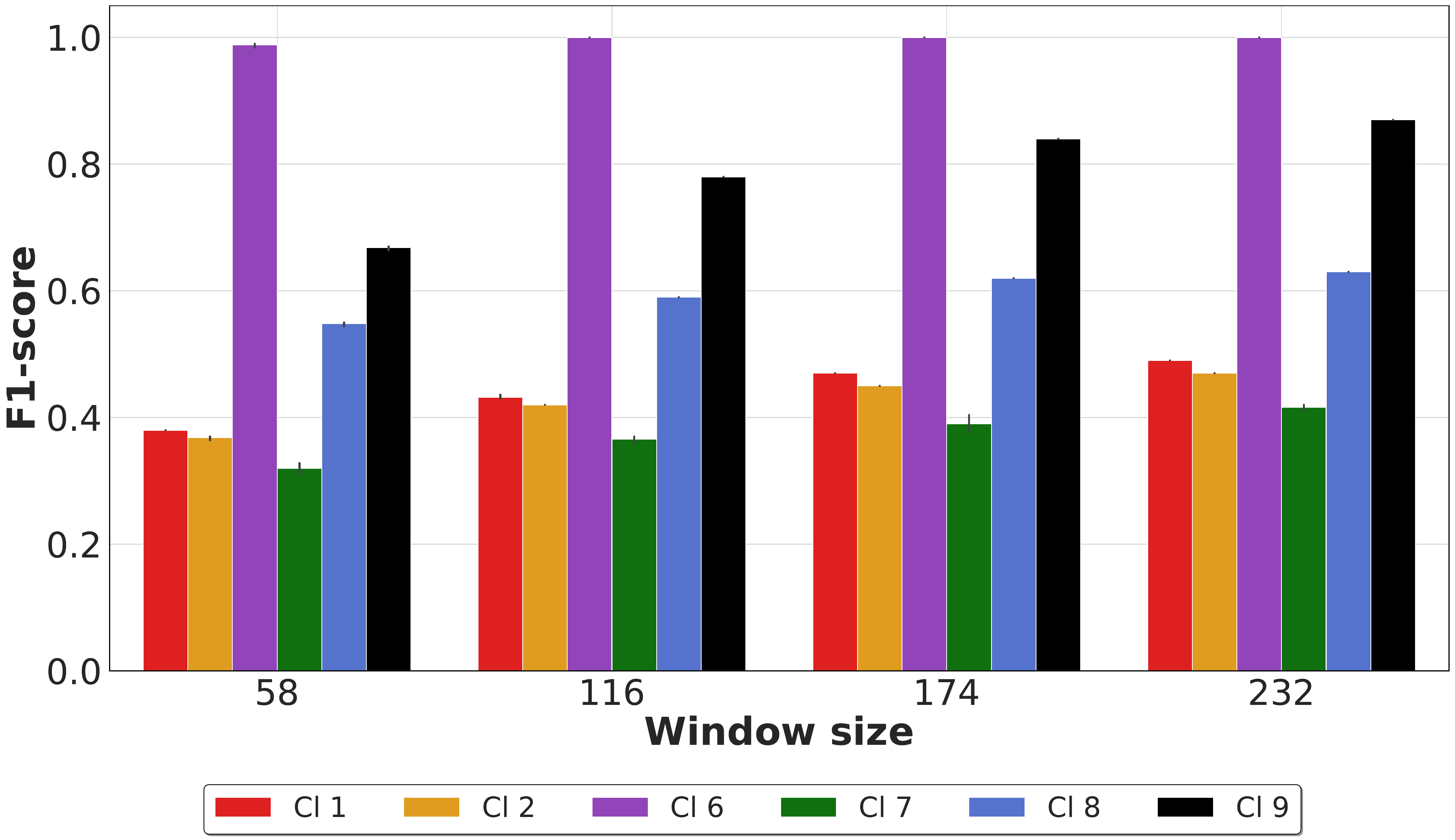}
\centering
\caption{Results obtained in the third configuration after the application of the four window sizes in the multiclass classification.}
\label{fig:multicassStatistics}
\end{figure}

This experiment creates a model where all the users are grouped, indicating to which user the signals belong. The three configurations of this experiment are the same as those implemented in the first experiment. Moreover, the framework uses the OneVsRest strategy, which consists in training a binary classifier for each of the ten subjects and comparing them. Regarding the third configuration, \figurename~\ref{fig:multicassStatistics} shows the results obtained after applying the four window sizes, which indicates that the performance improves when increasing the window size, similar to binary classification. The features used are all the statistics values obtained, that is, six statistical values multiplied by eight channels, making a total of 48 features.

\begin{figure}[ht]
\includegraphics[width=\columnwidth]{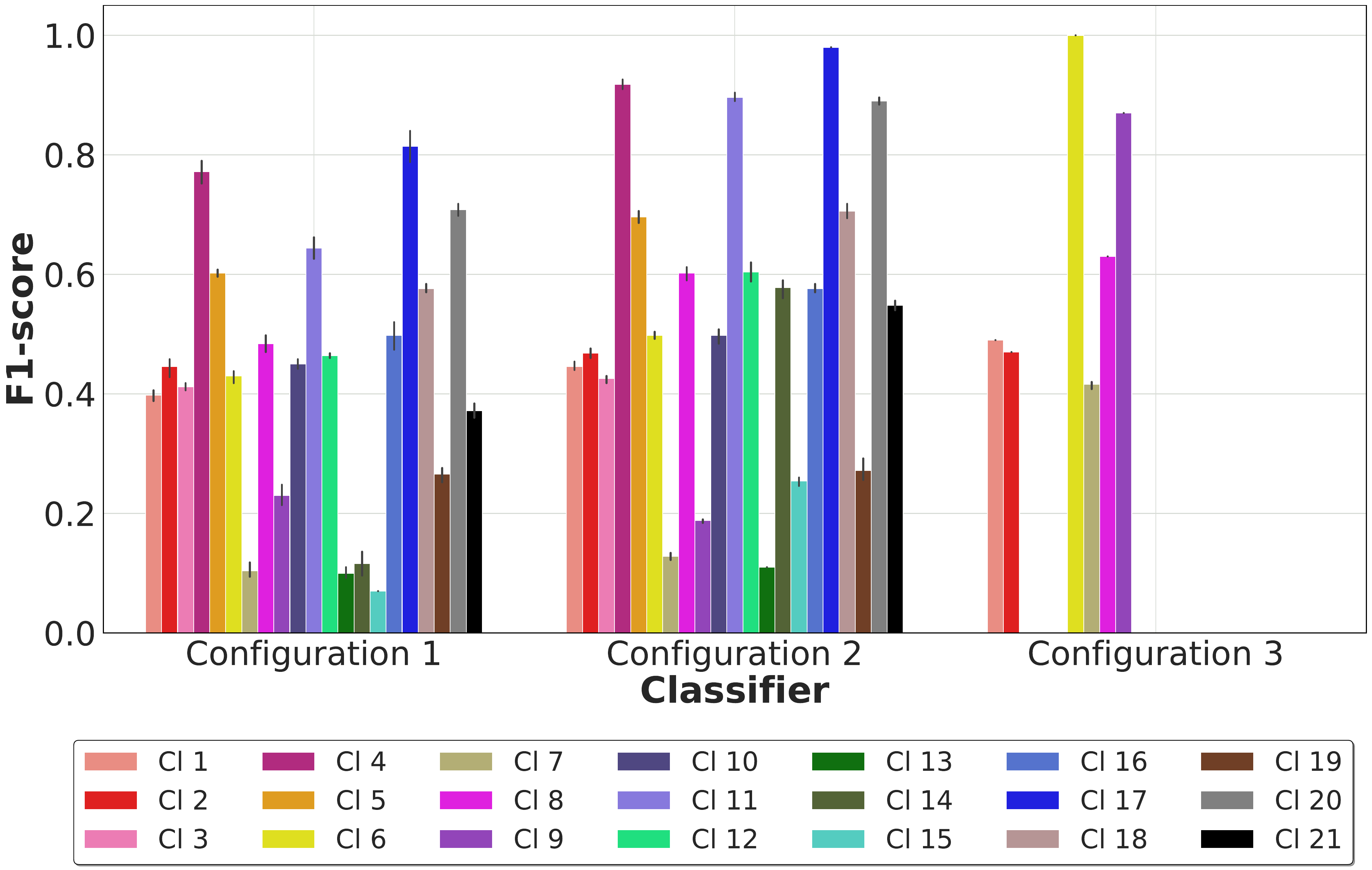}
\centering
\caption{Results obtained from the three configurations after performing a multiclass classification.}
\label{fig:multiclass}
\end{figure}

As in the previous experiment, \figurename~\ref{fig:multiclass} compares the results of the studied configurations. In the binary scenario, a performance close to 0.5 f1-score is similar to tossing a coin since the classifiers distinguish whether the data belong to that subject. In the multiclass scenario, a value close to 0.5 is a much better result as the classifier now determines which of the ten subjects the data belongs. In this sense, in the first configuration, it can be seen that several classifiers obtain results above 0.6, as is the case of classifiers 4, 5, 11, 17, and 20, with classifier 17 being the best and offering 0.81 f1-score. However, classifiers 7, 13, 14, and 15 obtain limited results, from 0.07 to 0.12.

In the second configuration, all the classifiers improve except classifier 9, which reduces from 0.23 to 0.19. Among the classifiers having a higher improvement, it is important to highlight classifier 4 (from 0.12 to 0.58 f1-score), classifier 11 (from 0.64 to 0.90), classifier 14 (from 0.12 to 0.58), and classifiers 17, 18, and 20 (0.98, 0.71 and 0.89 respectively). These results highlight a considerable improvement since there are four classifiers capable of distinguishing one user from the rest more than 89\% of the time, with classifier 17 being almost infallible.

As in binary classification, the third setup is performed on six classifiers from the total set of 21. The classification using statistical values improves the results obtained compared to employing epochs. In this sense, an improvement is seen in all the classifiers except in the second, which maintains its results. For example, classifier 6 goes from 0.5 using epochs to 1.00 with statistics. However, classifier 9 shows the most significant increase in performance, going from 0.19 to 0.87, which means that it has moved from being an uninteresting classifier to one capable of authenticating subjects with a high success rate.

\begin{figure}[ht]
\includegraphics[width=\columnwidth]{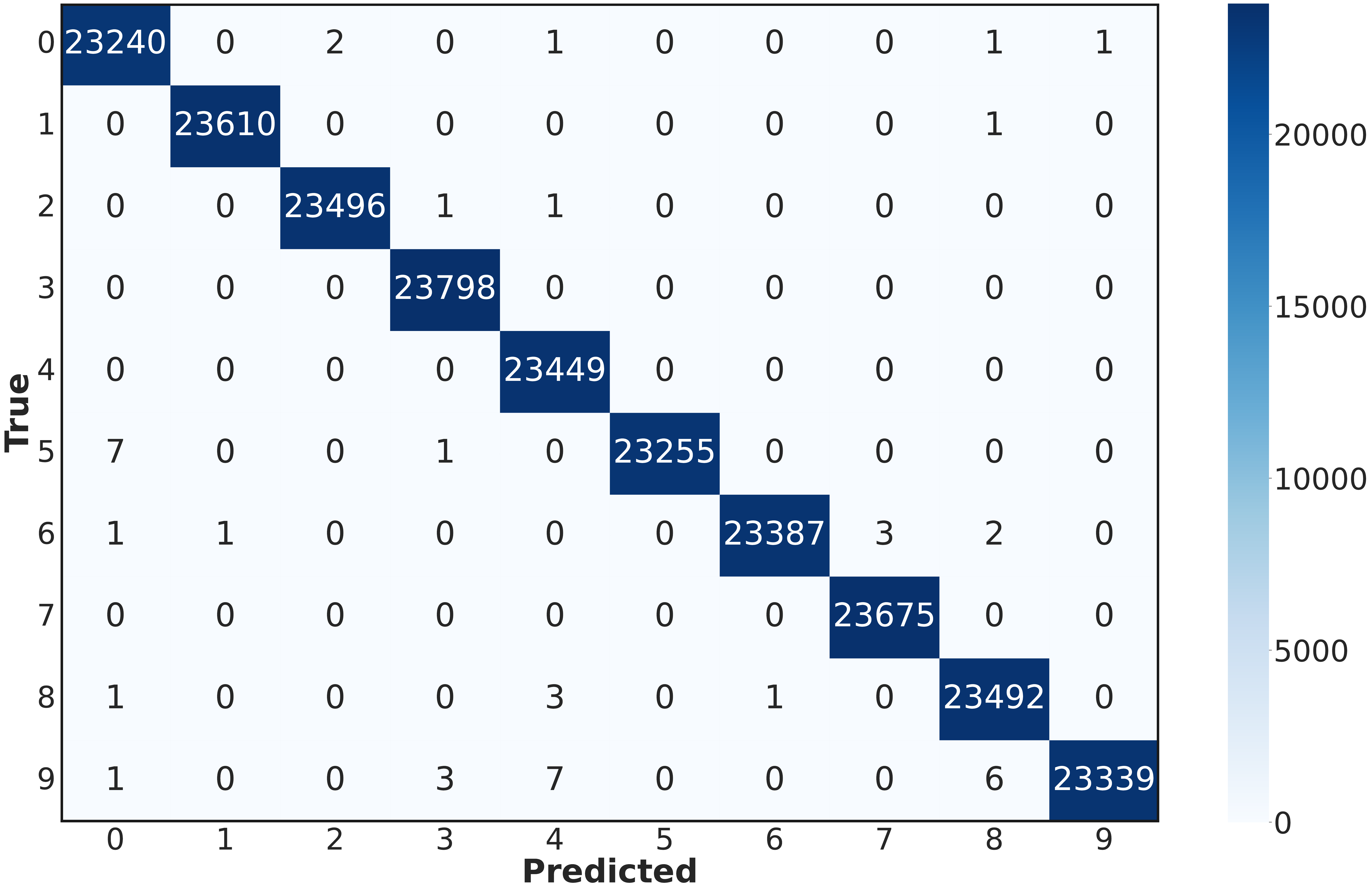}
\centering
\caption{Confusion matrix obtained for the best classifier used in multiclass classification (classifier 6).}
\label{fig:confusionMatrixMulticlass}
\end{figure}

Finally, \figurename~\ref{fig:confusionMatrixMulticlass} shows the confusion matrix obtained in the best classifier (classifier 6) after performing the multiclass classification using statistical values. Particularly, the training dataset contains 939 144 vectors, while the test dataset has 234 786 vectors. Additionally, the training process requires two hours and 22 minutes, using 1.13 Gb as storage, while evaluating the model takes 48 seconds.

\section{Discussion}
\label{sec:discussion}

The results presented in the previous section indicate that using additional processing techniques to the base case improves the performance of classifiers, although some techniques were ineffective during the experimentation. For instance, the inclusion of processing techniques such as the Notch filter or ICA, almost unexplored in the P300-based authentication literature, offers a considerable improvement in classification performance. Moreover, the results also conclude that using statistical values instead of epochs improves the performance of the authentication process in both experiments, as it describes the EEG signals better. In this context, this work has verified the hypothesis that better results can be obtained by augmenting the window size in the statistics experiment, resulting in a broader set of values that are more representative. Finally, both binary and multiclass ML classification approaches provide promising results, indicating the possibility of authenticating a user against the rest. They also identify to which user the brain waves tested belong, which is quite helpful in a corporate scenario where different use cases can arise.

\tablename~\ref{table:comparationResults} compares this work with the literature analyzed in Section~\ref{sec:related}. Their differences in terms of experimentation parameters, the processing techniques used, and the classification approaches followed were already presented in \tablename~\ref{table:literature}. Thus, this section presents a comparison in terms of the algorithms used and their performance. First, it is crucial to highlight that the literature analyzed indicates the performance based on accuracy instead of f1-score. Because of that, the values for this work have also been expressed using accuracy. Concerning these results, \tablename~\ref{table:comparationResults} indicates that the proposed solution presents the highest accuracy possible for classification approaches, being close to 100\%. 

\begin{table}[!htb]
\centering
\caption{Comparison of the results obtained in this work and the works studied in the literature based on accuracy.}
\label{table:comparationResults}
\begin{tabular}{@{}lccc@{}}
\toprule
Reference & Year & Algorithms & Accuracy\\
\midrule
\cite{Gupta:auth_BCI:2012} & 2012 & LDA & 90\%
\\ \midrule

\cite{Yu:auth_BCI:2014} & 2014 & FLDA & 83.1\%
\\ \midrule

\cite{Koike-akino:auth_BCI:2016} & 2016 & LDA, PLS & 96.7\%
\\ \midrule

\cite{Zeng:auth_BCI:2019} & 2019 & HDCA-GA & 94.26\%
\\ \midrule

\cite{Kaongoen:auth_BCI:2020} & 2020 & FLDA & \begin{tabular}{@{}c@{}}0 FRR \\ 0.003 FAR \end{tabular}
\\ \midrule

\cite{Rathi:auth_BCI:2021} & 2021 & QDA & 97.0\% \\ \midrule

This work & 2023 & RF & \begin{tabular}{@{}c@{}} Binary: $\approx$100\% \\ Multiclass: $\approx$100\% \end{tabular} \\\midrule
\end{tabular}
\end{table}

\section{Conclusion}
\label{sec:conclusions}

This work proposes the design and implementation of a framework able to authenticate users based on EEG signals. Particularly, it uses the Oddball paradigm to elicit P300 potentials as a cerebral response to known visual stimuli within a set of unknown stimuli. This publication defines an experimental scenario where ten subjects wear a non-invasive BCI while visualizing images. Moreover, the framework implements different signal processing and ML classification techniques since the literature lacks a comprehensive analysis of these dimensions.
 
The performance of the framework is evaluated based on two different experiments. The first experiment uses a binary classification approach to authenticate a subject against the rest, while the second employs multiclass classification to identify to which user the signals belong. Furthermore, three experimental configurations are defined for both experiments, incrementally testing different processing and classification strategies. The first configuration includes the Notch and Butterworth filters to remove noise from the signal, where the classification is performed using epochs. The second configuration adds the ICA filter to eliminate signal artifacts, also employing epochs. Finally, the third configuration inherits all three processing techniques but classifies utilizing statistical values extracted from the epochs, also using various sliding window sizes of vectors.

Focusing on the results obtained, both binary and multiclass ML classification approaches have proven helpful in authenticating subjects, surpassing the literature in performance. Regarding the previous configurations, there is a performance improvement when using all three processing techniques. Moreover, classifying with statistical values offer better results compared to epochs. Thus, the third configuration presents the best results, being random forest the algorithm offering the best performance for both experiments, close to 100\% f1-score. Furthermore, increasing the window size had a positive impact on the classification performance. 

In a nutshell, this publication advances the knowledge of EEG-based authentication, addressing a task not extensively studied in the literature. This paper also sheds light on the limited consensus from the literature in terms of signals processing for authentication. Moreover, this work demonstrates that both binary and multiclass authentication approaches are effective alternatives, indicating that employing statistical values of the epochs with a larger window size offers the best results.

In future work, the capture of the EEG signal could be improved by adding a more significant number of electrodes to have broader coverage of the brain. In the same way, other types of classifiers with an adaptation of the algorithms or an anomaly detector could be explored. Additionally, it would be interesting to increase the number of subjects to have more data variability. Finally, future work could explore different tasks to compare their performance for authentication. 

\section*{Acknowledgment}

This work has been partially supported by \textit{(a)} the Swiss Federal Office for Defense Procurement (armasuisse) with the CyberTracer and RESERVE projects (CYD-C-2020003), \textit{(b)} the University of Zürich UZH, and \textit{c)} Bit \& Brain Technologies S.L. under the project CyberBrain: Cybersecurity in BCI for Advanced Driver Assistance, associated with the University of Murcia (Spain).

\bibliography{references}{}
\bibliographystyle{plain}

\begin{IEEEbiography}[{\includegraphics[width=1in,clip]{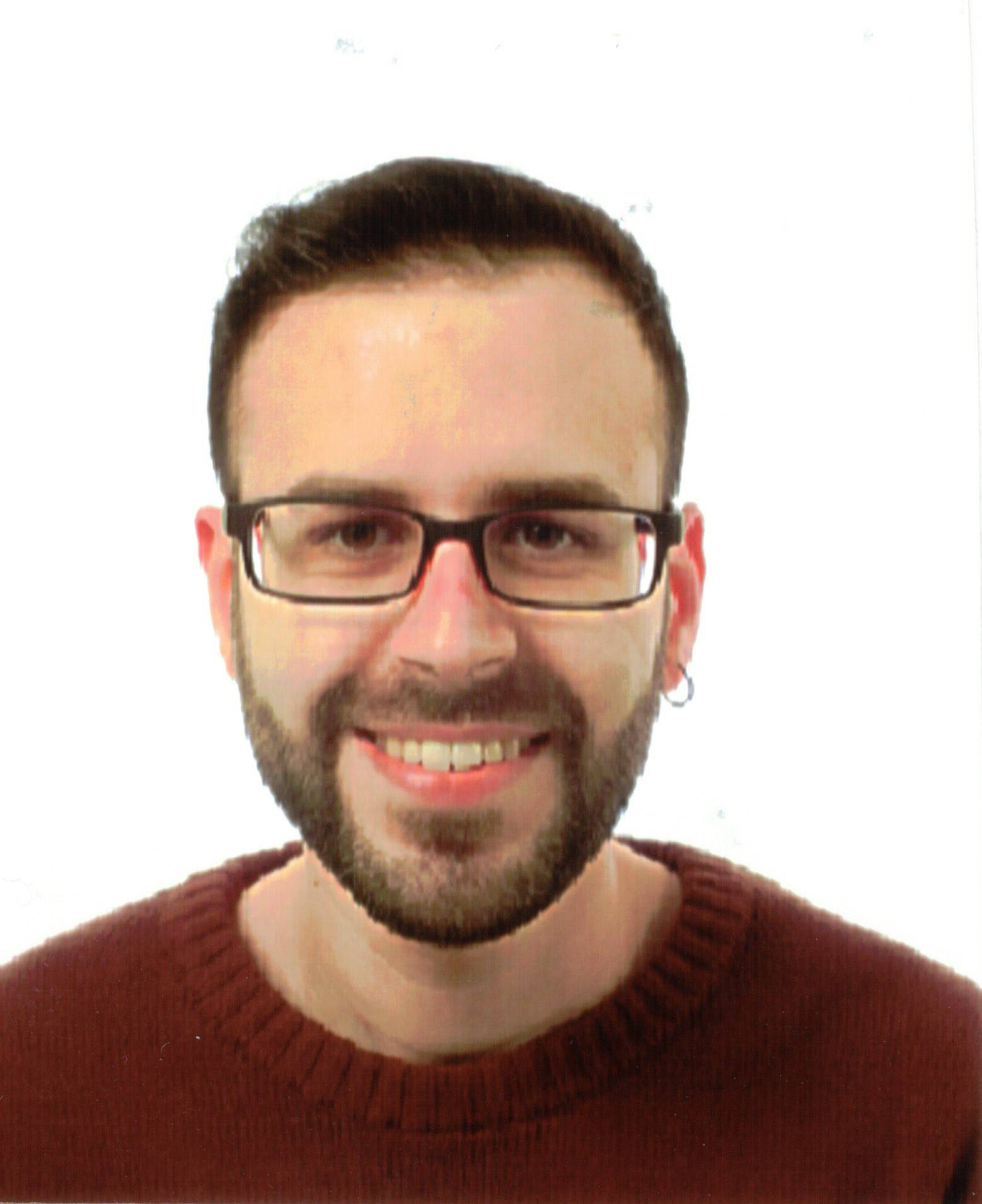}}]{Eduardo López Bernal} is a M.Sc. Student in Big Data at the University of Murcia. He received the B.Sc. Degree in Computer Science from the University of Murcia. His research interests focus on artificial intelligence and medical applications. Contact him at eduardo.lopez5@um.es
\end{IEEEbiography}

\begin{IEEEbiography}[{\includegraphics[width=1in,clip]{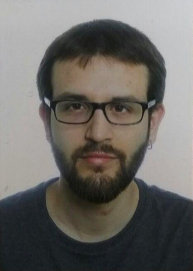}}]{Sergio López Bernal} is senior researcher in the Department of Information and Communications Engineering at the University of Murcia, Spain. He received the B.Sc. M.Sc. and Ph.D. degrees in computer science from the University of Murcia, and the M.Sc. degree in architecture and engineering for the IoT from IMT Atlantique, France. His research interests include cybersecurity on brain-computer interfaces and network and information security. Contact him at slopez@um.es

\end{IEEEbiography}

\begin{IEEEbiography}[{\includegraphics[width=1in,clip]{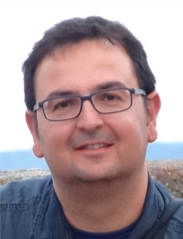}}]{Gregorio Martínez Pérez} is Full Professor in the Department of Information and Communications Engineering of the University of Murcia, Spain. His scientific activity is mainly devoted to cybersecurity and networking, where he has published 160+ papers. Contact him at gregorio@um.es
\end{IEEEbiography}


\begin{IEEEbiography}[{\includegraphics[width=1in,clip]{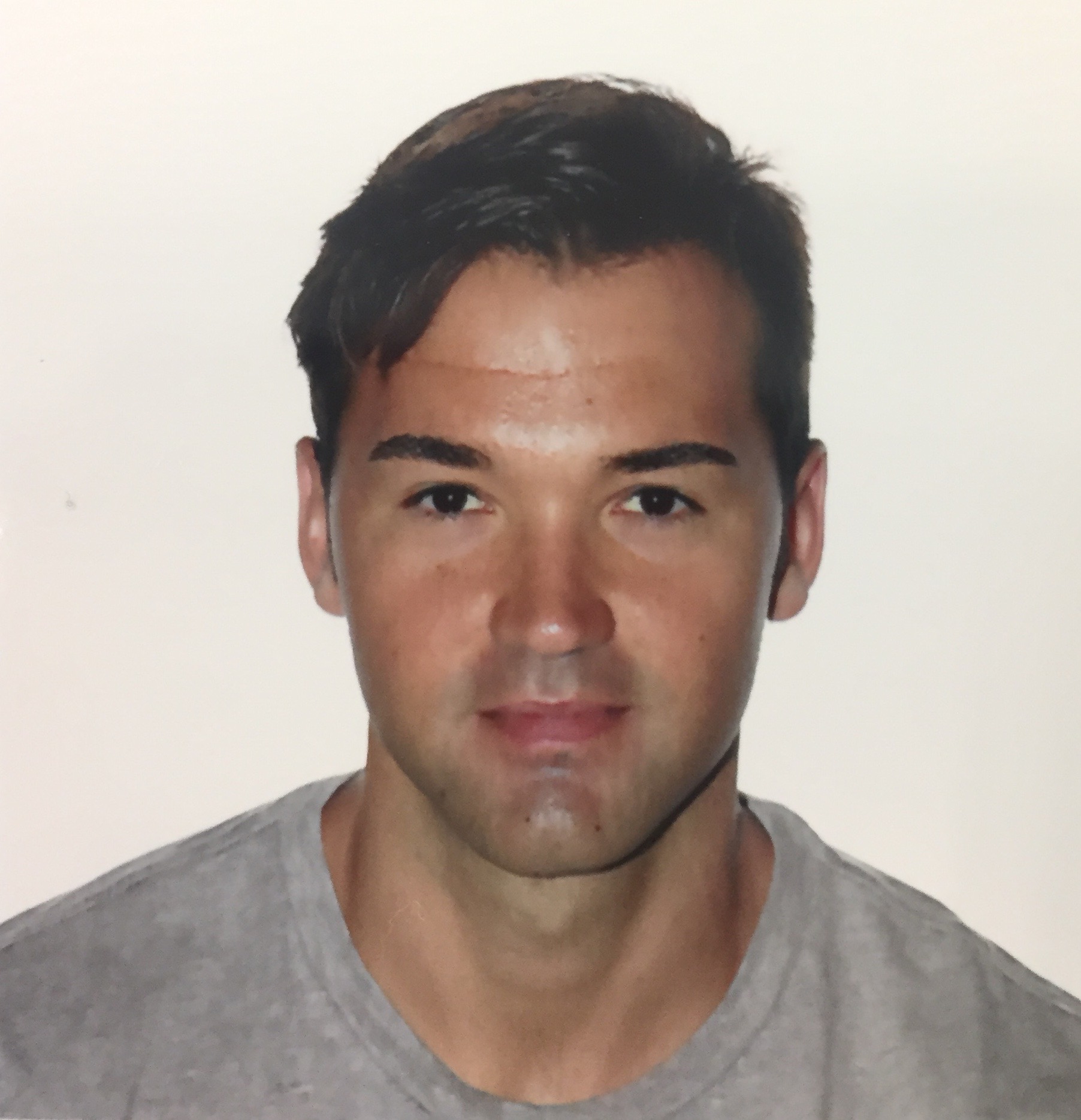}}]{Alberto Huertas Celdrán} is senior researcher at the Communication Systems Group CSG, Department of Informatics IfI, University of Zurich UZH. He received the MSc and PhD degrees in Computer Science from the University of Murcia, Spain. His scientific interests include cybersecurity, machine and deep learning, continuous authentication, and computer networks. Contact him at huertas@ifi.uzh.ch
\end{IEEEbiography}

\end{document}